\documentclass[iop]{emulateapj}

\usepackage{color,graphicx} 
\usepackage{amsmath,amssymb} 
\usepackage{tabularx}

\newcommand{\iso}[2]{\hbox{${}^{#1}{\rm #2}$}} 
\newcommand{\Msun}{\ensuremath{{M}_{\sun}}}

\shorttitle{AGB stars and the early solar system} \shortauthors{Wasserburg, Karakas \& Lugaro}

\begin{document}


\title{Intermediate-mass asymptotic giant branch stars and sources of $^{26}$Al,
$^{60}$Fe, $^{107}$Pd, and $^{182}$Hf in the solar system}


\author{G.~J. Wasserburg\altaffilmark{1,5}, 
Amanda I. Karakas\altaffilmark{2,3} and Maria 
Lugaro\altaffilmark{4,2}} \email{amanda.karakas@monash.edu}


\altaffiltext{1}{Lunatic Asylum, California Institute of Technology, Pasadena CA, 91125}
\altaffiltext{2}{Monash Centre for Astrophysics, School of Physics and Astronomy,
Monash University, VIC 3800, Australia}
\altaffiltext{3}{Research School of Astronomy and Astrophysics,
  Australian National University, Canberra, ACT 2611, Australia}
\altaffiltext{4}{Konkoly Observatory, Research Centre for Astronomy and Earth Sciences,
Hungarian Academy of Sciences, H-1121 Budapest, Hungary}
\altaffiltext{5}{{\em Deceased}}


\begin{abstract}

We explore the possibility that the short-lived radionuclides \iso{26}Al, \iso{60}Fe, 
\iso{107}Pd, and \iso{182}Hf inferred to be present in the
proto-solar cloud originated from $3-8\Msun$ Asymptotic Giant Branch (AGB) 
stars.  Models of AGB stars with initial mass above 5$\Msun$ are prolific producers of 
\iso{26}Al owing to hot bottom burning (HBB).  In contrast,
\iso{60}Fe, \iso{107}Pd, and \iso{182}Hf are produced by  
neutron captures: \iso{107}Pd and \iso{182}Hf in models $\lesssim
5\Msun$; and \iso{60}Fe in  models with higher mass.
We mix stellar yields from solar-metallicity AGB models into a cloud of
solar mass and composition to investigate if it is possible to explain
the abundances of the four radioactive nuclides at the Sun's 
birth using one single value of the mixing ratio between the AGB
yields and the initial cloud material. We find that AGB stars that
experience efficient HBB ($\geq 6$ \Msun) cannot provide a solution because they
produce too little \iso{182}Hf and \iso{107}Pd relative to \iso{26}Al
and \iso{60}Fe. Lower-mass AGB cannot provide a solution
because they produce too little \iso{26}Al relative to \iso{107}Pd and
\iso{182}Hf. A self-consistent solution may be found for AGB stars
with masses in-between ($4-5.5\Msun$), provided HBB is stronger than in 
our models and the \iso{13}C($\alpha$, n)\iso{16}O neutron source is mildly activated. 
If stars of M $< 5.5\Msun$ are the source of the radioactive nuclides,
then some  basis for their existence in proto-solar clouds needs to be
explored, given that the stellar lifetimes are longer than the
molecular cloud lifetimes.

\end{abstract}


\keywords{nucleosynthesis, abundances --- stars: AGB and post-AGB
--- ISM: abundances}



\section{Introduction}

A self-consistent solution for the origin of the inventory of short-lived radioactive 
nuclides inferred to be present in the early solar system from meteoritic analysis 
is still missing. Proposed solutions include core collapse 
supernovae \citep[e.g.,][]{takigawa08,pan12} as well as low and intermediate-mass Asymptotic 
Giant Branch (AGB) stars \citep[e.g.,][]{wasserburg06}. Interestingly, a few isotopes 
(e.g., \iso{53}Mn) can only be synthesized via explosive nucleosynthesis and are not produced in AGB 
stars. Some isotopes such as \iso{10}Be, \iso{26}Al, \iso{36}Cl, \iso{41}Ca, and \iso{53}Mn can also be 
produced by spallation reactions induced by Galactic and solar cosmic rays \citep{gounelle06}. 
Notably, a stellar source is favored for \iso{26}Al 
\citep[e.g.][]{duprat07,fitoussi08}.

When considering the results from core-collapse supernovae (SNeII) as possible 
contributors to the inventory of short lived nuclei, we note 
the following: 1) The ratio \iso{26}Al/\iso{27}Al in these sources is
not very high, with production typically $\sim 5 \times 10^{-3}$ \citep{rauscher02,lugaro14b}; 
2) the ratio of \iso{60}Fe/\iso{56}Fe predicted is $\sim 2.4 \times
10^{-3}$; 3) \iso{53}Mn is abundantly produced, where the ratio
\iso{53}Mn/\iso{55}Mn $\approx  0.15$. This is not very different from
the earlier results of \citet{woosley95}.  As noted in 
\citet{wasserburg06} these results require dilution factors of $\approx
10^{-2}$ to $10^{-4}$ between the SNeII yields and the proto-solar
cloud in order to account for the proto-solar ratios of \iso{26}Al/\iso{27}Al,
\iso{60}Fe/\iso{56}Fe, and \iso{53}Mn/\iso{55}Mn in the 
early solar system \citep[see, e.g., Fig~S1 of][]{lugaro14b}. 
It follows that SNeII cannot explain the \iso{26}Al inventory nor can
they significantly contribute to the Fe and Mn isotopes.

The emphasis here is on AGB production of 
the four short lived nuclei with mean-lives less than about 
10$^{7}$~yrs. The list of 
isotopes include \iso{26}Al (with a mean life $\overline{\tau}_{26}$ = 1.03 Myr), 
\iso{60}Fe ($\overline{\tau}_{60}$ = 3.75 Myr), \iso{107}Pd 
($\overline{\tau}_{107}$ = 9.38 Myr), and 
\iso{182}Hf ($\overline{\tau}_{182}$ = 12.8 Myr). These isotopes
can be produced in AGB stars by proton captures (\iso{26}Al) or by
neutron captures (\iso{60}Fe, \iso{107}Pd, \iso{182}Hf).

The isotope \iso{26}Al is a by-product of the MgAl chain 
operating in hydrogen burning environments \citep[e.g.,][]{arnould99}. Intermediate-mass AGB 
stars that experience hot bottom burning (HBB) can produce \iso{26}Al in copious quantities 
\citep{mowlavi00,karakas03a,izzard07,ventura11b}. HBB occurs when the temperature at the base 
of the convective envelope exceeds $50\times 10^{6}$~K, hot enough for proton capture 
nucleosynthesis \citep{bloecker91,lattanzio92,boothroyd92}.  HBB changes the surface 
composition because the whole convective envelope is constantly mixed into the hot 
region, with a mixing time of the order of $\approx 1$~year. The
minimum stellar mass for HBB to occur 
depends on the initial composition as well as the input physics used in the calculations 
\citep{ventura05a,ventura05b}. For solar metallicity, which we define here to be $Z=0.014$ 
adopting the solar composition of \citet{asplund09}, the minimum mass for HBB in our models is 
4.5$\Msun$ \citep{karakas14b}. Note that \iso{26}Al is easily destroyed by (n,$\alpha$) and 
(n,p) reactions so it cannot be produced by neutron captures.

Charged particle reactions on isotopes heavier than Si are unlikely to occur at AGB 
temperatures \citep{iliadis16}. For this reason the heavier radioactive nuclides \iso{60}Fe, 
\iso{107}Pd, \iso{182}Hf can be synthesized in AGB stars only by neutron captures occurring in 
the He-rich shell. 
While \iso{60}Fe is predominantly produced by neutron captures 
occurring in massive stars \citep{limongi06}, it can also be made in intermediate-mass 
AGB stars \citep{trigo09,lugaro12}. 
For the isotopes heavier than Fe, \iso{107}Pd and \iso{182}Hf, the main 
processes of neutron-capture 
nucleosynthesis are the $slow$ 
neutron-capture process and the $rapid$ neutron-capture process \citep[the $s$ and the $r$ process, 
respectively;][]{meyer94,kaeppeler11}. The $s$ process has been confirmed observationally 
to operate in low-mass AGB stars \citep{gallino98,abia02} and is a possible source 
of both \iso{107}Pd and \iso{182}Hf \citep{lugaro14b}.

Previously, the $r$ process was considered the dominant site of \iso{182}Hf in the Galaxy, however,
\citet{lugaro14b} 
pointed out that there is a good basis for the production of \iso{182}Hf in AGB stars since 
the lifetime of the precursor nucleus \iso{181}Hf in stellar environments is not too short:
The nuclear structure of the \iso{181}Hf nucleus used by \citet{takahashi87} was the basis of 
the decrease of the half life of \iso{181}Hf from $\simeq$ 42 days to $\simeq$ 3 hours 
in stellar interiors and the attribution 
of the origin of \iso{182}Hf to the $r$ process. However, due to new data 
on the states of \iso{181}Hf by \citet{bondarenko02} the decrease in
the half-life is now minimal.   This 
permits the inclusion of \iso{182}Hf in the inventory of AGB products and not the result 
of multiple $r$ process events as inferred by \citet{wasserburg94} from comparison with the 
abundance of \iso{129}I ($\overline{\tau}_{129}$ = 22.6 Myr), which can only be produced by 
the $r$ process. 
In the report by \citet{lugaro14b}, updated and revised models are presented together with an 
extensive discussion of the ratios \iso{107}Pd/\iso{108}Pd and \iso{182}Hf/\iso{180}Hf for a 
wide range of stellar masses. A time of 10-30 Myr from the last AGB $s$-process 
event was obtained to match the \iso{107}Pd/\iso{108}Pd and 
\iso{182}Hf/\iso{180}Hf ratios in the early solar system, 
during which the \iso{26}Al/\iso{27}Al produced by this intermediate-mass star would have 
completely decayed.  A separate \iso{26}Al source was assumed and no discussion was given in 
relation to the other short lived isotope \iso{60}Fe.
Here, we follow in detail 
the possible implications of the important revision on the AGB production of \iso{182}Hf 
to the scenario of an 
AGB source for some short-lived nuclei.

We present a detailed analysis of the possibility that
the isotopic shifts in the solar system for the four radioactive nuclei considered here
were due to 
injection of freshly synthesized radioactive nuclei, using the latest set of AGB star yields 
from \citet{karakas16}. We begin with a brief overview of AGB nucleosynthesis relevant to the 
production of the short lived nuclides found in the early solar system (Sec.~\ref{sec:agb}).
In Sec.~\ref{sec:model} we consider the extent to which any self-consistent solution for the 
estimated solar inventory 
can be found for the relative masses of the fresh stellar ejecta to the mass of the 
proto-solar cloud. A key to the dilution factor is the abundance ratio of short lived 
nuclei relative to stable isotopes of the same element in the AGB ejecta and the ratios at 
some reference time in the early solar system. There are 
reliable data estimating the 
abundance ratios at some times in the early solar system for \iso{26}Al/\iso{27}Al
(which we further discuss in Appendix~\ref{app:cai}),
\iso{107}Pd/\iso{108}Pd and \iso{182}Hf/\iso{180}Hf, but not for \iso{60}Fe/\iso{56}Fe, as we 
discuss in Appendix~\ref{app:fe60initial}. For completeness, 
in Sec.~\ref{sec:limitations} we discuss the potential issues with current AGB models 
and their impact on our results. In Sec.~\ref{sec:conclude} we present our conclusions.

\section{AGB star nucleosynthesis} \label{sec:agb}

Low and intermediate-mass stars cover a range in mass from 0.8 -- 8$\Msun$ for solar 
metallicity \citep[see Fig. 1 from][]{karakas14dawes}. Nucleosynthesis during the AGB is 
driven by He-shell instabilities. These thermal pulses (TP) may result in mixing between the 
H-exhausted core and the envelope; this is known as third dredge up (TDU), which alters the 
composition of the envelope by bringing the products of He-shell burning and the elements 
produced by the $s$-process to the stellar surface. For a 
review of AGB stars and their associated nucleosynthesis we refer to \citet*{busso99}, 
\citet{herwig05}, and \citet{karakas14dawes}.

Low-mass AGB stars with initial masses $M \lesssim 4\Msun$ have their surface compositions 
altered primarily by TDU, which results in enrichments in carbon, nitrogen, fluorine, and 
$s$-process elements 
\citep{busso01,abia02,karakas07b,karakas10a,cristallo11,cristallo15,karakas16}. In comparison, 
intermediate-mass AGB stars with initial masses $M\gtrsim 4\Msun$ experience the second dredge-up 
during the early AGB and HBB during the thermally pulsing AGB \citep[e.g.,][]{ventura13}. The 
surface chemistry of intermediate-mass stars therefore shows the results of proton-capture 
nucleosynthesis, with some contribution from the He-shell depending on the amount of TDU 
\citep{karakas12,ventura13,fishlock14b,cristallo15}.

The AGB models we are using in this study are from \citet{karakas16}. 
In brief, we use the stellar structure from detailed stellar evolution calculations
as input into a post-processing code that calculates the abundance changes due to 
nuclear reactions and mixing. We use 328 isotopes from the neutron to polonium
and roughly 2500 reactions from the JINA database as of May 2012. We refer to
\citet{karakas16} for further details on the numerical method and the input physics
used in the calculations. 

In \citet{karakas16} we compare our results to other AGB models in the 
literature including the models of \citet{cristallo15} and \citet{pignatari16}, while 
\citet{ventura15} compared intermediate-mass AGB models with HBB from \citet{karakas10a} and 
\citet{ventura13}. The summary is that the low-mass ($<$ 4 \Msun) models from 
\citet{cristallo15} are comparable in terms of their nucleosynthesis to the low-mass models 
from \citet{karakas16}, especially for heavy elements produced by the $s$ process. In 
contrast, the higher-mass models of \citet{karakas16} experience HBB at much higher 
temperature at a given mass compared to the models by \citet{cristallo15}, and also show much 
deeper TDU. Models by \citet{pignatari16} are comparable to the models by 
\citet{karakas16} for intermediate-masses, in terms of the depth of TDU and HBB temperatures 
\citep[see also models by][]{weiss09,colibri}. Models by \citet{ventura13} show even higher 
HBB temperatures than those by \citet{karakas16} for 
the same mass and composition but have much 
less TDU. The implications of these differences for the radio-nuclei discussed 
here and our results are detailed in Sec~\ref{sec:limitations}.

\begin{figure}
    \begin{center}
     \includegraphics[width=0.95\columnwidth]{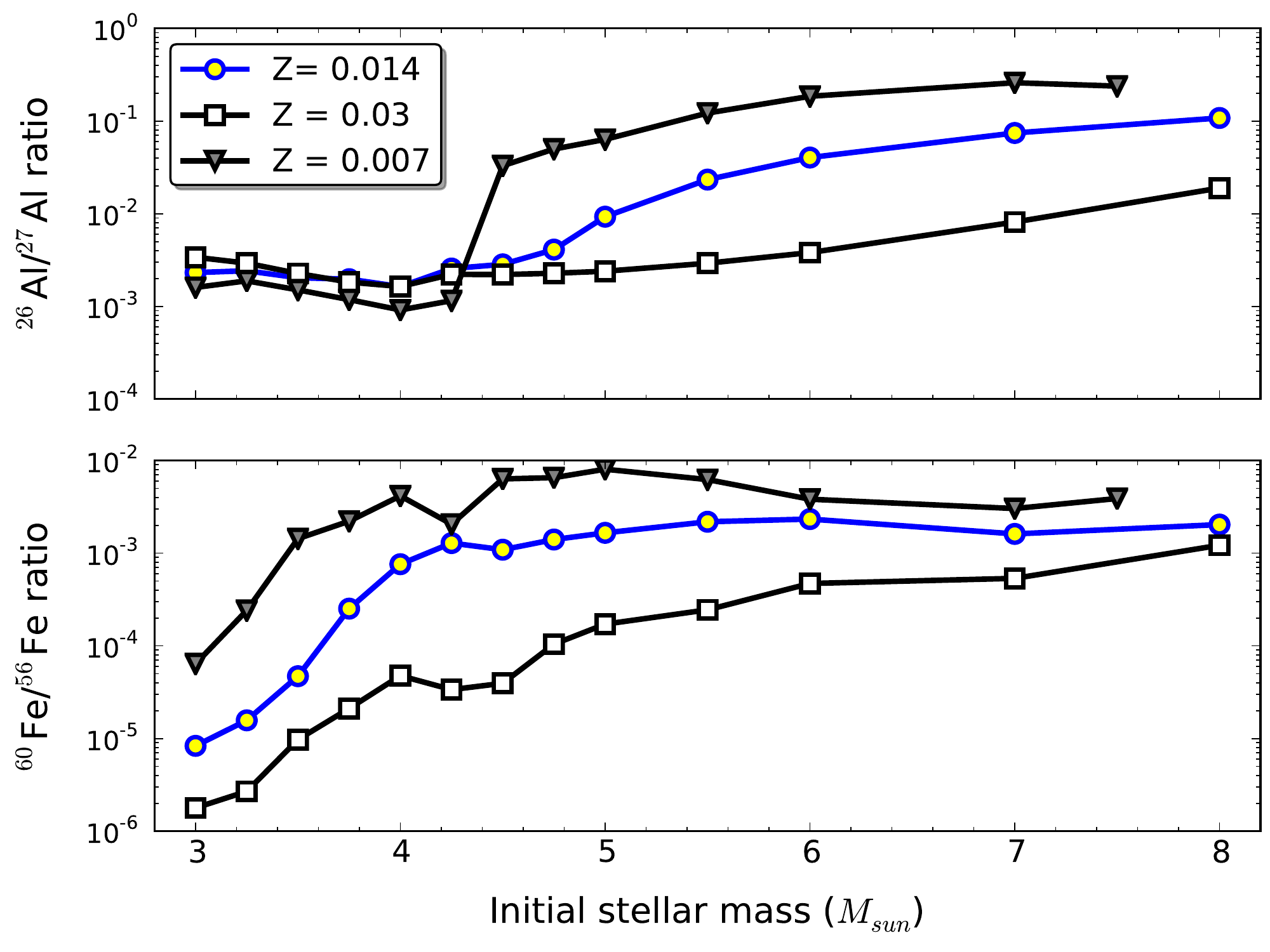} \caption{Predicted isotopic ratios for 
\iso{26}Al/\iso{27}Al and \iso{60}Fe/\iso{56}Fe (ratios are shown by number) as a function of 
initial stellar mass $M\ge 3\Msun$, for the three metallicities included in 
\citet{karakas16}. The ratios are calculated from the surface composition after the final 
thermal pulse. These are almost the same as the 
ratios calculated from the stellar yields because the yields 
are determined when most of the mass is lost from the star and this is near the tip of the 
AGB.}
    \label{fig1}
  \end{center} \end{figure}

In Fig.~\ref{fig1} we show the predicted \iso{26}Al/\iso{27}Al and \iso{60}Fe/\iso{56}Fe 
ratios for the models with initial mass 
$M \ge 3\Msun$ using data from \citet{karakas16}. The initial ratios are 
zero. From this figure we can see that the major difference between low-mass ($1.5-4\Msun$) AGB stars 
and intermediate-mass stars is the production of \iso{26}Al. HBB
results in copious \iso{26}Al production, with ratios $\approx 0.1$, 
in contrast to the situation for C-rich lower-mass stars which
generally have ratios $ < 10^{-2}$ \citep[e.g., see also][]{vanraai08}.

\begin{figure}
    \begin{center}
     \includegraphics[width=0.95\columnwidth]{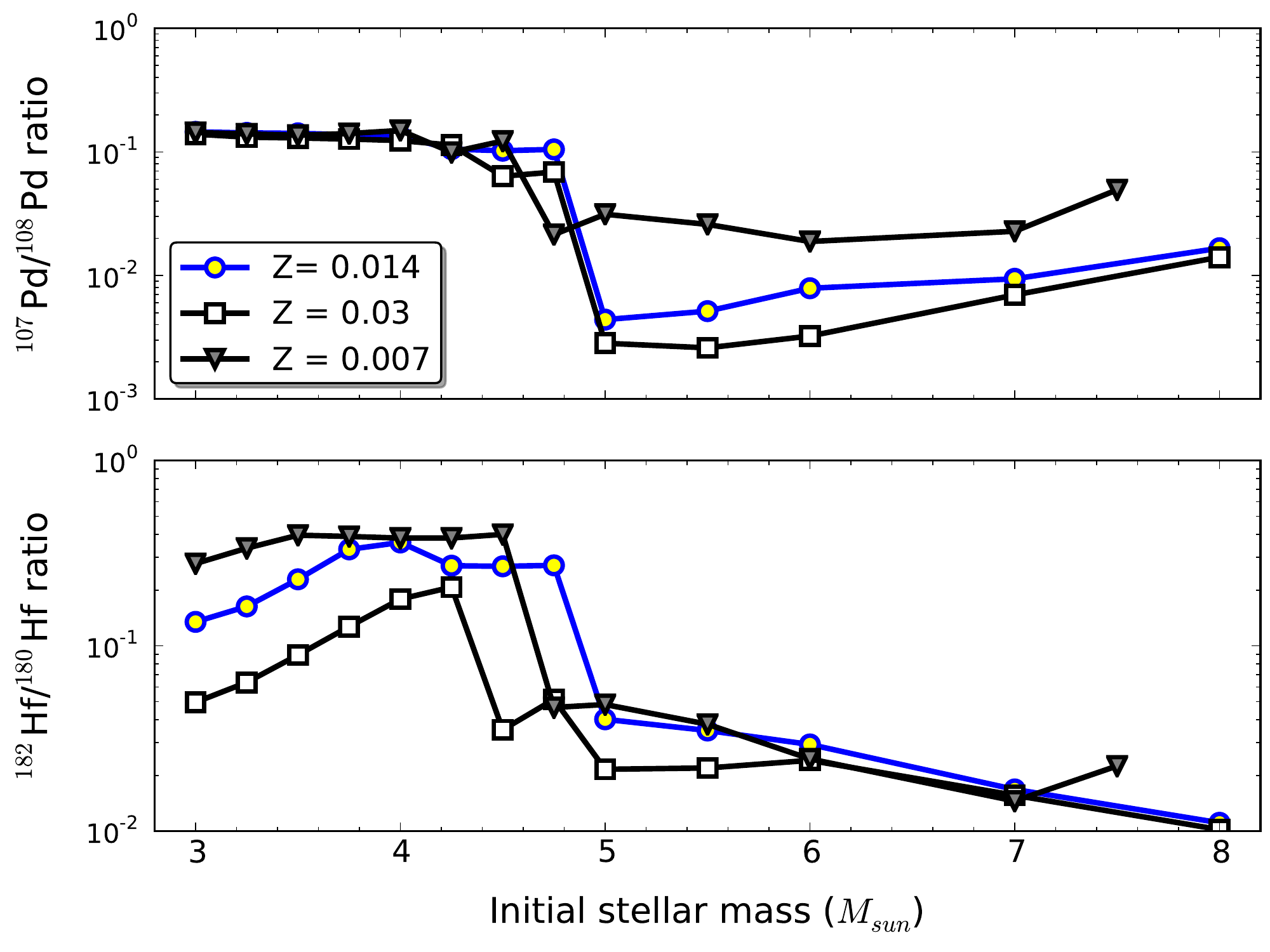} \caption{Same as Fig.~\ref{fig1} 
except for the \iso{107}Pd/\iso{108}Pd and \iso{182}Hf/\iso{180}Hf ratios.}
    \label{fig2}
  \end{center} \end{figure}

The minimum \iso{26}Al/\iso{27}Al ratio required in the envelope of an AGB star
is $\approx 2\times 10^{-2}$  in order to produce enough \iso{26}Al to
explain the amount inferred present in the early solar system \citep{wasserburg06}. 
From Fig.~\ref{fig1} we see that only models with masses above $4.5\Msun$ 
satisfy this criterion.  If we are to consider a lower mass star of $\approx 3\Msun$ as being 
responsible for the inventory of radioactive nuclides, we need to invoke some form of slow 
non-convective transport mechanism to explain the \iso{26}Al.
Such deep mixing is invoked to occur in the envelopes of low-mass ($\lesssim 2\Msun$) red 
giant branch stars \citep[e.g.,][]{gilroy89,gilroy91}. Evidence comes from observations of 
lower \iso{12}C/\iso{13}C and C/N ratios compared to theoretical models 
\citep{charbonnel94,boothroyd95,nollett03,charbonnel07a,eggleton08}. This process results in 
proton captures producing \iso{13}C and \iso{14}N. 
If it occurs also in AGB stars and if deeper layers are reached where the 
temperature is higher, then \iso{26}Al and \iso{17}O can also be produced
\citep[e.g.,][]{palmerini11}.

The mechanism responsible for the deep mixing is not known although in recent years 
parameterized versions of thermohaline mixing have been found to work, at least for the C and 
N isotopes in red giant branch stars \citep[e.g.,][]{angelou12}. Note
that observational evidence for deep mixing for elements heavier than
nitrogen is not well established from stellar spectra. 
Evidence for heavier isotopes instead comes from pre-solar grains, which are believed to have 
condensed in the atmospheres of evolved stars \citep[see extensive report by][]{zinner14}. 
However, no a priori prediction of the \iso{26}Al yield for low-mass AGB
stars is possible to be used for dilution calculations. Instead, the degree of deep mixing required to give 
the observed \iso{26}Al/\iso{27}Al ratio is calculated to match the other observations. In contrast, 
for models with HBB the \iso{26}Al yields are directly calculated for a stellar model. This 
is a direct result of the elevated temperatures in these more massive systems.

\subsection{The $s$ process in AGB stars} 

The isotopes \iso{60}Fe, \iso{107}Pd and \iso{182}Hf are produced exclusively by 
neutron-capture reactions. 
The main neutron source in low-mass AGB stars of $M \lesssim 4\Msun$ is the 
\iso{13}C($\alpha$,n)\iso{16}O reaction \citep{abia01,abia02}. CN cycling does not leave enough 
\iso{13}C nuclei in the He-intershell to produce enough $s$-process elements to match 
observations \citep{busso01}.  The solution to this problem is to assume that some partial 
mixing occurs between the H-rich envelope and the intershell at the deepest extent of each 
TDU. The protons are captured by \iso{12}C to produce a region rich in \iso{13}C, known as a 
\iso{13}C ``pocket''. The inclusion of \iso{13}C pockets in theoretical calculations of AGB 
stars is one of the most significant uncertainties affecting predictions of the $s$ process 
\citep[see discussions in][]{busso99,herwig05,kaeppeler11,karakas14dawes}. 

The details of how we include \iso{13}C pockets in our models is discussed 
in \citet{karakas16}. Briefly, at the deepest extent of each TDU
episode we include protons into the top layers of the
He-rich intershell region. Those protons are quickly captured by the
abundant \iso{12}C and converted into \iso{13}C and \iso{14}N by CN
cycle reactions.  \citet{fishlock14b} compared the shape and size of
the \iso{13}C pockets from this method to those calculated more
self-consistently by \citet{cristallo11} and found good agreement. For
models $M \le 3\Msun$ we include protons down to a depth in mass in
the He-intershell of $2\times 10^{-3}\Msun$, which results in a
\iso{13}C pocket that is $\approx 1/10$ of the mass of the
He-intershell.

In intermediate-mass stars the He-intershell becomes hot enough to activate the 
\iso{22}Ne($\alpha$,n)\iso{25}Mg reaction inside the TP. For masses in the transition between 
mild and strong HBB (4-5$\Msun$ for solar metallicity) there will be a contribution from both the 
\iso{13}C and the \iso{22}Ne neutron source. In intermediate-mass AGB stars with strong HBB 
($M \gtrsim 5\Msun$), evidence suggests that \iso{13}C pockets do not form and 
the $s$-process is the result of the \iso{22}Ne reaction \citep{goriely04,garcia13}. In the 
$Z=0.014$ models from \citet{karakas16} we include \iso{13}C pockets in models $< 5\Msun$, 
with the size of the \iso{13}C pocket decreasing as a function of
increasing stellar mass.
We also test the case of including \iso{13}C pockets in the 5$\Msun$ model.
Because the intershell region is smaller by roughly an order of magnitude in this
case we reduce the mass over which we mix protons by a similar factor to
$1\times 10^{-4}\Msun$ \citep[e.g., as discussed in][]{karakas16}. 

The predicted ratios from stellar models are shown in Figs.~\ref{fig1} and~\ref{fig2}.
The \iso{60}Fe/\iso{56}Fe ratio follows \iso{26}Al/\iso{27}Al where intermediate-mass 
AGB stars over 5$\Msun$ produce the most \iso{26}Al and \iso{60}Fe. The reason is that to 
produce \iso{60}Fe it is necessary to bypass the branching point at \iso{59}Fe 
($\overline{\tau}_{59}$ = 64 days), which requires neutron densities above $\sim 10^{9}$ 
n/cm${^3}$.  Such high neutron densities can only be produced inside
TPs when the temperatures and densities are high enough to
activate the \iso{22}Ne neutron source, above $300\times 10^{6}$~K.
This is achieved inside models of intermediate mass.

In contrast, the ratios of \iso{107}Pd/\iso{108}Pd and \iso{182}Hf/\iso{180}Hf are
relatively flat for models $ < 5\Msun$ but drop by an order of
magnitude in the more massive AGB stars. The reason is 
that significant amounts of these isotopes can only be synthesized if the neutron
exposure is relatively high, which is when the \iso{13}C pocket is included in the
low-mass models, which allows for activation of the \iso{13}C($\alpha$, n)\iso{16}O
neutron source reaction. Hence, high absolute abundances in 
the He-rich region (and consequently a strong signature at the stellar surface) 
are possible only when the \iso{13}C pocket is included.  The isotope \iso{182}Hf is 
further dependent on the branching point at \iso{181}Hf, which has similar mean-life as 
\iso{59}Fe, hence its abundance reaches a maximum in models of $\simeq$ 4 \Msun, where both 
the \iso{13}C and \iso{22}Ne neutron sources are activated.
As noted above, in intermediate-mass AGB stars the mass of the
He-intershell drops by an order of magnitude. While these models are
predicted to experience many more TPs than their lower mass counterparts 
\citep[e.g.,][]{doherty14a} the total amount of dredged-up material is
lower or similar to their lower mass counterparts \citep[see Fig.~1 from][]{karakas16}.

\section{The mixing model} \label{sec:model}

The mixing model used here represents the addition of freshly synthesized nuclei to the solar nebula 
in the framework of a molecular cloud with a variety of stars and the consideration of the 
times of formation of objects in the early solar system. Relative to some time ($\tau_{0}$) 
in the very early solar system, debris from an AGB star that underwent major mass loss at a (negative) 
time $\tau_{\rm AGB}$ is mixed with one \Msun\ of 
matter of solar composition with the mixing factor $F= M_{\rm AGB}/(M_{\rm AGB}+ \Msun) \approx M_{\rm AGB}/\Msun$.
Here $M_{\rm AGB}$ represents the debris from the AGB star
and is a small fraction of the total mass lost from the AGB  star's envelope.
We use exactly the same formalism described in detail in \citet{wasserburg06} (see their Eqs.~6 and 7).
For each isotope pair $i$ (unstable), $j$ (stable) listed in Table~\ref{table1} we define $F_{i,j}$
as the mixing factor derived
by imposing that the mixing produces the ratios $R_{i,j}$ observed in the
early solar system:
$$
F_{i,j}=\frac{R_{i,j}}{R^{\rm AGB}_{i,j} \times PF^{\rm AGB}_{j}},
$$
where $R^{\rm AGB}_{i,j}$ is the isotopic ratio from the AGB stellar yields and $PF^{\rm AGB}_{j}$
is the AGB production factor of the stable isotope $j$, relative to its initial solar abundance.
Clearly, a self-consistent
solutions for all the four isotope pairs considered here needs to produce the same value for the
four $F_{i,j}=F$.

\subsection{Input to the model}

The reference data used for all 
our calculations are given in Tables~\ref{table1},~\ref{table2a} and~\ref{table2b}. We use the 
stellar model results of \citet{karakas16} for $Z=0.014$ and proto-solar abundances from 
\citet{asplund09}.
Table~\ref{table1} shows the mean lifetime of species $i$, $\overline{\tau}_{i}$, given in 
years, the ratios \iso{26}Al/\iso{27}Al, \iso{107}Pd/\iso{108}Pd, \iso{182}Hf/\iso{180}Hf at 
the Calcium-aluminum (CAI) reference time with \iso{26}Al/\iso{27}Al
$=5.5\times 10^{-5}$ in the early solar system. 
The ratio of \iso{60}Fe/\iso{56}Fe is further discussed in Appendix~\ref{app:fe60initial}. 
Tables~\ref{table2a} 
and~\ref{table2b} show the predicted ratios of \iso{26}Al/\iso{27}Al, \iso{60}Fe/\iso{56}Fe, 
\iso{107}Pd/\iso{108}Pd and \iso{182}Hf/\iso{180}Hf from the AGB yields calculated by 
\citet{karakas16}. Table~\ref{table2a} shows the predictions for intermediate-mass AGB models 
which do not include a \iso{13}C pocket. Table~\ref{table2a} shows predictions for two masses 
(3$\Msun$ and 5$\Msun$), which include \iso{13}C pockets.

One further complication is related to the timescale of the formation of the objects 
from whose analysis the initial abundance in the solar system is derived.
At time $\tau_{0}$, CAIs are formed; at later times  ($\tau_{\rm  P1}$) 
proto-planet formation occurs with a variety of types of chemical fractionation (Fe-Ni, 
FeS, silicate separation from bulk material with major chemical fractionation); this is 
followed at later times ($\tau_{\rm P2}$) by cooling of planetary material and the freezing 
in of chemical fractionation and diffusion. Some of the data on meteoritic samples are made on 
different chemical phases in a single object to produce an internal isochron. The time this 
represents is when the object cooled ($\tau_{\rm P2}$), 
not necessarily when it formed ($\tau_{\rm P1}$) and 
gives the ratio (of say \iso{107}Pd/\iso{108}Pd) in that object at $\tau_{\rm P2}$. CAIs
typically contain clear evidence of \iso{26}Al with a maximum 
value of \iso{26}Al/\iso{27}Al $=5.5 \times 10^{-5}$. These CAIs are used to represent the 
initial reference time ($\tau_{0}$). \iso{26}Al is used because of the short mean life 
($\overline{\tau}_{\rm 26Al} = 1.03\times 10^{6}$~yr) and its widespread nature in CAIs. CAIs 
are surmised to be condensates from a mass of hot solar nebular gas. The actual mechanism 
which produced CAIs is not in fact known, nor do we know that they were produced at one time 
or place or at what stage of growth the Sun had attained. It is known that CAI formation took 
place over an extended time ($> 10^{5}$ yr) \citep{hsu00}. More discussion can be found in 
Appendix~\ref{app:cai}.

The key short lived isotopes discussed here are \iso{26}Al, \iso{60}Fe, 
\iso{182}Hf and \iso{107}Pd. Of these, only the values at CAI formation 
time for \iso{26}Al and \iso{182}Hf are well determined. The thorough 
and extensive study by \citet{burkhardt08} and \citet{kruijer13} have established internal 
isochrons for Hf-W on CAIs. This gives a direct comparison for these nuclei of refractory 
elements at what is plausibly the same time.  An insightful and thorough investigation of 
\iso{182}Hf/\iso{180}Hf in bulk FeNi meteorites was carried out by \citet{kruijer14b} 
corrected for cosmic ray effects using \iso{196}Pt as a monitor. These workers established 
initial values of \iso{182}Hf/\iso{180}Hf for Fe-Ni segregation from silicates. These results 
are not internal isochrons but represent the times when bulk Hf-W chemical fractionation took 
place between metal and silicate masses in parent planets. These workers find that there was a 
rather short time between $\tau_{\rm P1}$ and $\tau_{\rm CAI}$ \citep[several million years, 
see][their supplemental data, Table 6]{kruijer14a}. In contrast, for \iso{107}Pd we know from 
internal isochrons for three meteorites (Gibeon, Duchesne, Muonionalusta) that 
\iso{107}Pd/\iso{108}Pd $= 2.4 \times 10^{-5}$ \citep{chen96,horan12} and see 
\citet{matthes15} for the most precise value for Muonionalusta. The \iso{107}Pd/\iso{108}Pd 
ratio for these samples is the value when the diffusion process stopped between the 
co-existing phases in these objects. It is some $\tau_{\rm P2}$. It is not the same time as 
that for bulk Fe-Ni-silicate segregation. \citet{matthes15} have the most precise and thorough 
analysis and discussion of the \iso{107}Pd-\iso{107}Ag system. 

\begin{table*}
\begin{center}
\caption{Isotopic Ratios in the early solar system at the CAI ($R^{0}_{i,j}$) 
and cooling ($R^{\tau_{P2}}_{i,j}$) reference times.}
\label{table1}
\begin{tabular}{l c c c} \hline \hline
Isotope & $\overline{\tau}_{i}$/yr & $R^{0}_{i,j}$ & $R^{\tau_{P2}}_{i,j}$ \\
\hline
\iso{26}Al  & $1.03\times 10^{6}$ & $5.5\times 10^{-5}$ & -- \\
\iso{60}Fe  & $3.75\times 10^{6}$ & $< 10^{-6}$ & $< 2\times 10^{-6}$ \\
\iso{107}Pd & $9.38\times 10^{6}$ &
$2.4\times10^{-5}\exp(\tau_{\rm P2}/\overline{\tau}_{i})$ & $2.4\times 10^{-5}$ \\
\iso{182}Hf$^{\rm a}$ & $12.8\times 10^{6}$ & $9.72\times 10^{-5}$ & -- \\
\hline \hline
\end{tabular}
\\
(a) Using data from \citet{burkhardt08}. Note that \citet{kruijer14b} give $(1.018 \pm 0.043) 
\times 10^{-4}$.
\medskip\\
\end{center}
\end{table*}

\begin{table*}
\begin{center}
\caption{Ratios by number in the net ejecta from \citet{karakas16} 
for models without \iso{13}C pockets.}
\label{table2a} 
\begin{tabular}{l c c c c} \hline \hline
Isotope & 5$\Msun$ & 6$\Msun$ & 7$\Msun$ & 8$\Msun$ \\
\hline
\iso{26}Al/\iso{27}Al & 9.47$\times 10^{-3}$ & 4.24$\times 10^{-2}$
& 7.29$\times 10^{-2}$ & 8.85$\times 10^{-2}$  \\
\iso{60}Fe/\iso{56}Fe   & 9.55$\times 10^{-4}$ & 1.14$\times 10^{-3}$ &
7.11$\times 10^{-4}$ & 7.45$\times 10^{-4}$ \\
\iso{107}Pd/\iso{108}Pd & 3.42$\times 10^{-3}$ & 5.37$\times 10^{-3}$ &
7.67$\times 10^{-3}$ & 1.19$\times 10^{-2}$   \\
\iso{182}Hf/\iso{180}Hf &  3.52$\times 10^{-2}$ & 2.24$\times 10^{-2}$ &
1.11$\times 10^{-2}$ & 5.47$\times 10^{-3}$   \\
\hline \hline
\end{tabular}
\medskip\\
\end{center}
\end{table*}

\begin{table*}
\begin{center}
\caption{Ratios by number in the net ejecta from \citet{karakas16}
for models with \iso{13}C pockets.}
\label{table2b}
\begin{tabular}{l c c} \hline \hline
Isotope & 3$\Msun^{\rm a}$ & 5$\Msun^{\rm b}$ \\
\hline
\iso{26}Al/\iso{27}Al & 2.28$\times 10^{-3}$ & 9.50$\times 10^{-3}$  \\
\iso{60}Fe/\iso{56}Fe & 6.74$\times 10^{-6}$ & 9.12$\times 10^{-4}$  \\
\iso{107}Pd/\iso{108}Pd & 1.45$\times 10^{-1}$ & 9.97$\times 10^{-2}$  \\
\iso{182}Hf/\iso{180}Hf & 1.25$\times 10^{-1}$ & 2.47$\times 10^{-1}$  \\
\hline \hline
\end{tabular}
\\
a) For the 3$\Msun$ model with a standard \iso{13}C pocket, see details
in \citet{karakas16}.\\
b) Using the one calculation of a 5$\Msun$ model with a \iso{13}C pocket from
\citet{karakas16}.
\medskip\\
\end{center}
\end{table*}

\begin{table*}
\begin{center}
\caption{Mixing ratios $F_{i,j}$ for the models without 
\iso{13}C pockets.}
\label{table3a}
\begin{tabular}{l c c c} \hline \hline
Mass & $F_{26,27}$ & $F_{107,108}$ & $F_{182,180}$ \\
\hline
5$\Msun$  & $5.8\times 10^{-3}$ & $7.0\times10^{-3}\exp(\tau_{\rm P2}/\overline{\tau}_{i})$
& $3.0\times 10^{-3}$  \\
6$\Msun$  & $1.3\times 10^{-3}$ & $4.5\times10^{-3}\exp(\tau_{\rm P2}/\overline{\tau}_{i})$
& $4.3\times 10^{-3}$  \\
7$\Msun$  & $7.5\times 10^{-4}$ & $3.1\times10^{-3}\exp(\tau_{\rm P2}/\overline{\tau}_{i})$
& $8.8\times 10^{-3}$ \\
8$\Msun$ & $6.2\times 10^{-4}$ & $2.0\times10^{-3}\exp(\tau_{\rm P2}/\overline{\tau}_{i})$
& $1.8\times 10^{-2}$ \\
\hline \hline
\end{tabular}
\medskip\\
\end{center}
\end{table*}

\begin{table*}
\begin{center}
\caption{Mixing ratios $F_{i,j}$ for models with \iso{13}C pockets.}
\label{table3b}
\begin{tabular}{l c c c} \hline \hline
Mass & $F_{26,27}$ & $F_{107,108}$ & $F_{182,180}$ \\
\hline
3$\Msun$  & $2.4\times 10^{-2}$ & $1.74\times10^{-4}\exp(\tau_{\rm P2}/\overline{\tau}_{i})$
& $7.8\times 10^{-4}$ \\
5$\Msun$  & $5.8\times 10^{-3}$ & $2.4\times10^{-4}\exp(\tau_{\rm P2}/\overline{\tau}_{i})$
& $3.9\times 10^{-4}$  \\
\hline \hline
\end{tabular}
\medskip\\
\end{center}
\end{table*}

\begin{table*}
\begin{center}
\caption{$R_{60,56}^{0}$ calculated from $F_{182,180}$.}
\label{table4}
\begin{tabular}{c c} \hline \hline
Mass & $R_{60,56}^{0}$ \\
\hline
\multicolumn{2}{c}{ Models calculated with a \iso{13}C pocket.} \\ \hline
3$\Msun$ & $5.3 \times 10^{-9}$ \\
5$\Msun$ & $3.6 \times 10^{-7}$ \\ \hline
\multicolumn{2}{c}{ Models calculated without \iso{13}C pockets.} \\ \hline
5$\Msun$ & $2.9 \times 10^{-6}$ \\
6$\Msun$ & $4.9 \times 10^{-6}$ \\
7$\Msun$ & $6.3 \times 10^{-6}$ \\
8$\Msun$ & $1.3 \times 10^{-5}$ \\
\hline \hline
\end{tabular}
\medskip\\
\end{center}
\end{table*}

\subsection{Results}

To gain some insight into the problem of self consistent models, we first consider mixing 
ratios for \iso{26}Al/\iso{27}Al and \iso{182}Hf/\iso{180}Hf. 
Table~\ref{table3a} 
shows the values of $F_{i,j}$ 
for the three isotopic pairs for which early solar system ratios have been determined,
using the reference values at CAI time given 
in Table~\ref{table1} and the ratios in the ejecta (Tables~\ref{table2a} and~\ref{table2b}) 
for different stellar masses. 
It can be seen that the mixing ratio is very high for \iso{26}Al at lower masses and then 
decreases drastically, reflecting the much higher temperatures accessible in more massive 
stars. In contrast \iso{182}Hf produced by neutron captures gives low $F_{182,180}$ values at lower 
masses and then rapidly increases to very high mixing ratios. The only apparent solution for 
this couplet is at $\approx 5.5\Msun$. Higher mass values are excluded for this isotopic pair. 
For \iso{107}Pd, it is seen that $F_{107,108}$ always exceeds $F_{26,27}$. If we seek 
to match only \iso{107}Pd and \iso{182}Hf, we find that $\tau_{\rm P2}$ should be $\approx 9 \times 
10^{6}$~yr for the 7$\Msun$ case. This value is reasonable. For the
8$\Msun$ $\tau_{\rm  P2}$ is $\approx 18 \times 10^{6}$~yr instead.
For these high masses all 
solutions that can match the initial solar values require very high mixing ratios
($>4 \times 10^{-3}$) to obtain the right amounts of \iso{182}Hf and \iso{107}Pd. This then would also 
require the \iso{26}Al that is co-produced to have significantly decayed.  
This requires consideration of an 
AGB event that precedes the initial formation of the solar system by several million years 
($\tau_{\rm AGB} \approx 3 \times10^{6}$ yr). 

Now, we consider the ratio of \iso{60}Fe/\iso{56}Fe that would occur for intermediate-mass 
stars if the mixing ratio for \iso{182}Hf/\iso{180}Hf were used for \iso{60}Fe. 
We see from Table~\ref{table4} that for all cases above about $5\Msun$ the \iso{60}Fe/\iso{56}Fe
ratio to be expected at CAI 
time is above $10^{-6}$. While the abundance of \iso{60}Fe is not well established (see 
Appendix~\ref{app:fe60initial}), it is 
clear that \iso{60}Fe/\iso{56}Fe $<10^{-6}$ is the upper bound possible at CAI time from all the 
data available.
It follows that any attempt to attribute the origin of both \iso{182}Hf and \iso{26}Al to an 
intermediate-mass star is
excluded from consideration of \iso{60}Fe. We note that \citet{lugaro14b} (see their Fig~S1) for a 
$6\Msun$ also found that possible self-consistent solutions with $F \sim 0.005$ would have much too 
high a value for \iso{60}Fe/\iso{56}Fe. 

For $3\Msun$ (Table~\ref{table3b}) we see that the \iso{182}Hf and \iso{107}Pd are essentially 
concordant if $\tau_{\rm P2} =14$~Myr. It is evident that \iso{26}Al is grossly under produced by a 
factor of 31. This is typical of all low-mass AGB stars as was long recognized. 
For a 3$\Msun$ star to produce enough \iso{26}Al and match \iso{182}Hf 
would require \iso{26}Al/\iso{27}Al $\approx 2\times 10^{-2}$ in the envelope.
If one assumes that deep mixing (from non-convective transport mechanism)
was in effect, from the extensive report of \citet*{nollett03} this would 
require penetration of a circulating mass to temperatures close to that of the H burning zone 
($\log T \approx 7.7$~K). This is the same as the circulation penetration required for some 
oxide grains of circumstellar condensates \citep[see][]{zinner14}. For 3$\Msun$ the production 
of \iso{60}Fe is very low and using the same dilution factor as for \iso{182}Hf, gives 
\iso{60}Fe/\iso{56}Fe= $5.52\times10^{-9}$, far below the upper bound cited above.

The deep mixing needed to produce \iso{26}Al is known to be 
required in the envelopes of low-mass ($\lesssim 2\Msun$) red giant branch stars as discussed 
in Sec.~\ref{sec:agb}. Observational evidence for extra mixing in the envelopes of 
intermediate-mass stars of of $\approx 3\Msun$ stars is less clear but could come from the 
high He/H and N/O ratios observed in Type I and bipolar planetary nebulae, which likely 
evolved from intermediate-mass progenitors $\ge 2\Msun$ \citep{corradi95,karakas09}. The extra 
mixing mechanism operating in the envelopes of intermediate-mass stars of $\approx 3\Msun$ 
stars is however unknown but could be the combination of 
thermohaline and rotation-induced mixing \citep[e.g.,][]{charbonnel10}.

There is an issue with regard to the production of \iso{26}Al for $4-5.5\Msun$ stars. These 
are transitional as they lie at the border between no HBB and intense HBB ($ M>6\Msun$). If 
some penetrative extra mixing process or a stronger HBB
could be operative at around $5\Msun$, then one might 
appeal to that mechanism to make the dilution factors 
compatible between \iso{26}Al and \iso{182}Hf, for which case
\iso{107}Pd will essentially agree with data. It is also possible that \iso{13}C 
pockets may be operative as an important neutron source (i.e., normal $s$ process). With 
regard to the low-mass case ($3-4\Msun$) it is clear that a self consistent solution for 
\iso{26}Al, \iso{182}Hf, \iso{107}Pd with some form of extra mixing may be possible and gross 
overproduction of \iso{60}Fe avoided, but would not explain the existence of a FUN CAI showing 
the initial presence of \iso{182}Hf but no \iso{26}Al \citep{holst13}. Furthermore, the 
problem remains as to how these lower mass intermediate-mass star with 
long evolutionary lifetimes could be in molecular clouds with lifetimes of $\approx 10^{8}$~yr 
and contribute to the cloud medium.

\section{Limitations on the AGB model calculations} \label{sec:limitations}

The conclusions drawn here are limited by uncertainties in the models for the yields of 
intermediate-mass AGB stars. It is well established that these stars undergo HBB \citep[see recent 
overview by][]{ventura11A}. However, the quantitative effect of HBB in stellar models is 
dependent on how convective mixing is implemented. For the mixing length method used in our 
models, the temperature at the base of the convective envelope increases with the value of the 
free mixing length parameter, $\alpha_{\rm MLT}$. Other mixing schemes produce different 
results; the Full Spectrum of Turbulence (FST) models used by \citet{ventura13} result in 
higher HBB temperatures than we obtain, while the models of \citet{cristallo15} present 
typically lower temperatures for the same mass and metallicity. We expect massive AGB stars 
to produce \iso{26}Al but we cannot accurately establish at which initial stellar mass HBB may 
actually start. A problem affecting the production of \iso{26}Al by HBB is that the rate of 
the destruction reaction \iso{26}Al$+$p is uncertain \citep{siess08}.
Thus, an accurate \iso{26}Al yield cannot be well established.

The yields of all species are affected by the mass-loss rate. This is because mass loss 
determines the AGB lifetime, hence the number of thermal pulses, as well as the duration of 
HBB. Faster mass loss for example, results in lower yields of \iso{26}Al because there is less 
time for HBB to operate, and lower yields of \iso{60}Fe and \iso{182}Hf, because there are 
fewer TPs and TDU events. In our models, we used the semi-empirical mass-loss 
prescription by \citet{vw93}. The production of species in the He intershell also 
depends on the TDU efficiency. This remains a debated uncertainty for intermediate-mass 
AGB models \citep{frost96,mowlavi99a,kalirai14}. Models of massive AGB stars that experience 
no or little dredge-up \citep[such as the FRUITY model for 6$\Msun$,][]{cristallo15} do not 
present large yields for either \iso{60}Fe and \iso{182}Hf.

While there are clearly some uncertainties, we feel that some conclusions appear clear. The 
production of the early solar system inventory of \iso{182}Hf from massive AGB stars is inevitably 
accompanied by production of \iso{60}Fe to levels above those inferred to have been present in 
the early solar system. The presence of a \iso{13}C pocket could change this result, since in this case the 
elements which are produced from Fe seeds (including e.g., \iso{180}Hf and \iso{108}Pd) yield 
high isotopic ratios (c.f., \iso{182}Hf/\iso{180}Hf, \iso{107}Pd/\iso{108}Pd) in the stellar 
envelope. The elements that are not greatly enhanced by an intrinsic $s$ process (e.g., Ti, 
Fe, Ni etc) do not produce high isotopic ratios in the envelope (compare 
\iso{107}Pd/\iso{108}Pd, \iso{182}Hf/\iso{180}Hf with \iso{60}Fe/\iso{56}Fe in 
Tables~\ref{table2a} and~\ref{table2b}). For a case with a \iso{13}C pocket, the production of 
\iso{60}Fe can be kept small and that of \iso{182}Hf can be large. However, \iso{13}C pockets 
are not expected to be present in AGB stars suffering HBB, both theoretically 
\citep{goriely04} and observationally \citep{garcia13}. This means that a decoupling of 
\iso{182}Hf from \iso{60}Fe also gives low \iso{26}Al. We see no means of producing 
\iso{182}Hf without high \iso{60}Fe/\iso{56}Fe, unless the current nuclear physics inputs 
(neutron-capture cross sections of \iso{59}Fe and \iso{60}Fe, the decay rate of \iso{59}Fe, or the 
rates of the \iso{22}Ne$+\alpha$ reactions) are extremely inaccurate.

Thus even considering the model uncertainties we do not find a possible self-consistent 
solution for the origin of \iso{26}Al, \iso{60}Fe, and \iso{182}Hf is the early solar system
for initial masses $>6\Msun$.

\section{Conclusions} \label{sec:conclude}

From consideration of the results obtained in the stellar models of \citet{karakas16} of 
intermediate-mass stars and comparing the output of stars ranging in mass from 4$\Msun$ to 
8$\Msun$, we conclude that the inventory of \iso{26}Al, \iso{182}Hf, 
\iso{107}Pd and \iso{60}Fe assumed for the early solar system cannot be explained by sources 
of mass $> 6\Msun$. There is a clear need to establish stricter \iso{60}Fe/\iso{56}Fe values at 
the times of CAI formation. Sources of lower mass ($4-5.5\Msun$), which are transitional in 
nature, may play a significant role. As HBB is not a dominant feature of these stars, it is 
possible the extra mixing processes that produce \iso{26}Al and the formation of \iso{13}C pockets 
may permit a possible solution. This would then be similar to models of $2-3\Msun$ AGB stars 
as sources. The objection to low-mass AGB stars as a source of short lived nuclei for the 
solar system is based on the long time scales for evolution to the AGB phase as compared to 
the lifetime of a molecular cloud ($\sim 10^{6}$-$10^{7}$ yr).  The evolutionary time scales for 
5.5$\Msun$ and 3$\Msun$ stars is $\sim$~77 Myr and 650~Myr, respectively. These stars require 
efficient extra mixing and would not violate the \iso{60}Fe bound.  It is not evident that the 
time scales for stellar evolution for such stars is short enough for their contribution to 
nucleosynthesis in molecular clouds. The association of more 
massive star formation within molecular clouds is evident from many observations of OB 
associations. The conclusions drawn here point to a difficulty for relating the formation of 
the solar system to such a cloud. One possible solution is that there are always many older 
stars present within a molecular cloud. These are not, in general, co-moving with the cloud 
but are passing through it by differential motion.  If we consider the volume density of main 
sequence stars of $\approx 1\Msun$ in the solar neighborhood to be $\sim 1$ star/parsec$^{3}$ 
and taking the size of a cloud to be 30 parsec, then the number of stars in the corresponding 
volume is $\sim 3 \times 10^{4}$. Using a Salpeter initial mass function this gives $\sim 
10^{3}$ 3$\Msun$ stars in the cloud. This suggests that along the spiral arms of the galaxy, 
where the gas is concentrated, longer lived, lower mass stars ($2-5\Msun$) have a reasonable 
probability of evolving to planetary nebulae and mixing with clouds leading to new star 
formation. A serious answer depends on the appropriate astration rate as a function of stellar 
mass and the volume density of stars $>1 \Msun$ in the spiral arm region where the solar 
system was hatched.

\acknowledgments

This work was supported by the Epsilon Foundation (GJW). One of us acknowledges the aid of 
Anneila Sacajawea Sargent who guided us to Nick Scoville who responded to the question: 
``Could low mass stars from earlier generations be in a molecular cloud?''. ML is a Momentum 
(``Lend\"ulet-2014'' Programme) project leader of the Hungarian
Academy of Sciences. AK and ML warmly thank Yongzhong Qian for his
help with revising the manuscript after the passing of Jerry, and to
the Referee for comments which have greatly helped to improve the paper.

\appendix

\section{Issues concerning \iso{26}Al and CAIs} \label{app:cai}

There are three matters concerning \iso{26}Al in the CAIs that require attention. The first of 
these is the presence of CAIs and ultra-refractory oxides with
\iso{26}Al/\iso{27}Al ratios ranging 
from $5\times 10^{-5}$ to values some decades below this \citep[c.f.,][]{makide11}. These 
samples also have low \iso{18}O/\iso{16}O and \iso{17}O/\iso{16}O ratios, but have 
\iso{17}O/\iso{18}O of the terrestrial value (e.g., \iso{16}O enriched). This oxygen effect in 
CAIs was first discovered by \citet{clayton73}. These workers also showed the presence of 
\iso{16}O depleted material in phases in the same CAIs. This was the result of alteration of O 
in these phases. Several of these phases also exhibit clear excesses on \iso{26}Mg from 
\iso{26}Al decay. Note that some of the phases in CAIs with ``oxygen'' alteration have the 
canonical \iso{26}Al/\iso{27}Al ratio. See recent summary by \citet{krot14} of the oxygen problem 
and references therein.

The \iso{16}O enriched oxygen found in phases in CAIs and ultra-refractories is currently 
believed to represent the actual solar inventory inferred from measurements of the solar wind 
by the GENESIS spacecraft \citep{mckeegan11}.  If the original solar 
inventory of \iso{26}Al/\iso{27}Al is $\approx 5.5\times 10^{-5}$, then the CAIs and 
ultra-refractory grains (such as Al$_{2}$O$_{3}$) which have ``solar'' oxygen and 
\iso{26}Al/\iso{27}Al ranging from $\sim 5\times 10^{-5}$ to very low values must reflect the 
passage of time from an initial state or incomplete mixing of stellar debris with no 
\iso{26}Al and with no other detectable nuclear effects \citep[see][]{makide11}. The proposal 
that this might result from the very late injection of \iso{26}Al into the solar nebula in 
which no \iso{26}Al was present has been proposed. This late injection scenario would require 
that no other nuclear effects would be added and does not explain the well defined upper bound 
of \iso{26}Al/\iso{27}Al$=5.5 \times 10^{-5}$.

Alternatively, the refractories with very low to no \iso{26}Al/\iso{27}Al could represent 
on-going infall from the local interstellar medium 
over a time scale of $\approx 3 \times 10^{6}$ yr and the 
solar oxygen then reflecting on-going infall from that medium. This long time scale view is in 
conflict with the typical collapse times of $\sim 10^{5}$ yr \citep[c.f.,][]{boss11}. However, 
it is well known that differential motion of an accreting star through a cloud over $3 \times 
10^{6}$ yr can readily provide the last $\sim 3$\% of a solar mass from ongoing infall due to 
gravitational sweep up \citep{hoyle39,bondi44,edgar04a,edgar04b}. It is thus reasonable that 
the range of \iso{26}Al/\iso{27}Al might be due to this process of on-going late infall from 
an initial homogeneous source region. This model also implies that the ultra-refractories and 
CAIs formed over an extended time period and that some had to form by shock heating of 
infalling debris.

The second issue is the multi-stage growth of CAIs. It is well known that individual CAIs 
($\sim 1$~cm) are a composite of different material.  \citet*{elgoresy85} showed that there 
are distinct multi-layers and \citet{hsu00} showed that layers in a single CAI represent 
differences of $\sim 10^{5}$~yr or more using \iso{26}Al as a chronometer.

The third and last issue relates to the problem of terrestrial type oxygen which dominates the 
``normal'' Fe, Mg-rich chondrules and the terrestrial planets so far sampled. The alteration of 
oxygen in CAIs \citep[c.f.,][]{krot14} and the origin of terrestrial type oxygen is a mystery 
that is much discussed and little understood by all parties. Many of the phase considered to 
be primary and have \iso{26}Al/\iso{27}Al $\approx 5\times 10^{-5}$ have undergone oxygen 
exchange by some unknown mechanisms.  With these caveats, we consider that the issue of 
possible stellar sources of \iso{26}Al as discussed here are sound.

\section{The problem of the initial \iso{60}Fe/\iso{56}Fe} \label{app:fe60initial}

As a guide, we note that the steady state ratio for the Galaxy based on gamma ray fluxes from 
the decay of \iso{60}Fe and \iso{26}Al are (\iso{60}Fe/\iso{56}Fe)$_{\rm GALS} = 1.5 \times 
10^{-7}$ and (\iso{26}Al/\iso{27}Al)$_{\rm GALS} = 1.0 \times 10^{-5}$ 
\citep{diehl10,diehl16}. There is no data on \iso{60}Fe that can be used from CAIs because: 1) 
Wide spread isotopic anomalies in both Fe and Ni in CAIs which prevent one from obtaining 
meaningful results on \iso{60}Ni; and 2) the Fe in CAIs is not, in general a primary 
constituent. Fe is not an ultra refractory element and the frequent occurrence of FeS in CAIs 
is interpreted to reflect late stage alteration processes that are known to have occurred. 
With regard to data obtained on planetary differentiates, to be of merit it must be connected 
to the initial \iso{26}Al inventory. For time scales $> 5$~ Myr, \iso{26}Al has decayed and 
any connection in time to CAIs is obscure. In any case, as the effects in \iso{60}Ni become 
exceedingly small, the problem of widespread isotopic heterogeneity in the solar system 
becomes severe.

The required datum is \iso{60}Fe/\iso{56}Fe at the time when \iso{26}Al/\iso{27}Al $\approx 5 
\times 10^{-5}$. In attempting to obtain some estimate of this it has been necessary to 
analyze Fe, Mg chondrules from unequilibrated ordinary chondrites (UOC). These chondrules are 
made of silicates with ``terrestrial'' type oxygen. Previous workers have shown that some of 
these chondrules contain Al-rich phases and exhibit excesses of \iso{26}Mg/\iso{24}Mg 
correlated with \iso{27}Al/\iso{24}Mg \citep{hutcheon89}. Such samples thus may exhibit clear 
evidence of \iso{26}Al and can be related to the CAIs by using the inferred 
\iso{26}Al/\iso{27}Al ratio as a measure of time. Measurements of \iso{60}Fe on samples of 
chondrules from unequilibrated chondrites (UOC) and bulk chondrites have yielded a wide range 
of results. It must be recognized that these measurements are exceedingly difficult. Precise 
measurements by \citet{tang15} give \iso{60}Fe/\iso{56}Fe $\approx 5 \times 10^{-9}$ at the 
time of crystallization of a chondrule from Semarkona (UEC) and an inferred initial value of 
$\sim 10^{-8}$. An investigation by \citet{tachibana06} gave \iso{60}Fe/\iso{56}Fe $\approx 
(2-4) \times 10^{-7}$ at the time of formation of some chondrules. However no evidence for the 
presence of \iso{26}Al was obtained in either report.  In the study by \citet{mishra14} 
measurements were of both Al-Mg and Fe-Ni isotopic systematics on chondrules from some UOC 
samples. Some of the Al-Mg data were obtained by \citet{rud08}. We restrict our attention to 
those samples with rather clear \iso{26}Mg/\iso{24}Mg -- \iso{27}Al/\iso{24}Mg correlations, 
defined \iso{26}Al/\iso{27}Al initial values, and with a reasonably justified correlation of 
\iso{60}Ni/\iso{62}Ni versus \iso{56}Fe/\iso{62}Ni. Using the \iso{26}Al data as a measure of 
time, five samples define a value of (\iso{60}Fe/\iso{56}Fe)$_{\rm CAI}$ in the range of $5 
\times 10^{-7}$ to $10^{-6}$ \citep{mishra14}. It is this data set which is the basis of the 
upper bound used here. There are no data available which indicate a higher value.

\bibliographystyle{apj} 
\bibliography{apj-jour,library,lib}


\end{document}